\documentclass[twocolumn]{article}
\usepackage{graphicx,verbatim}
\usepackage{graphicx}
\usepackage{cite}
\usepackage{array}
\usepackage{amsmath}
\usepackage{hyperref}
\usepackage{geometry}
\usepackage{amsmath}
\usepackage{amssymb}
\usepackage[table,xcdraw]{xcolor}
\usepackage{booktabs}
\usepackage{pgfplots}
\pgfplotsset{compat=1.18}
\usepgfplotslibrary{groupplots}
\usepgfplotslibrary{polar}
\usepackage{tabularx}
\usepackage{adjustbox}
\usepackage{subcaption} 
\usepackage{adjustbox}
\usepackage{makecell}

\newcommand{\method}{UTS}
\newcommand{\pipeline}{UTS}
\newcommand{\net}{L-ViT}

\title{\textbf{U}nit-Based Histopathology \textbf{T}issue \textbf{S}egmentation  via Multi-Level Feature Representation}

\author{
Ashkan Shakarami\textsuperscript{1}\thanks{Corresponding author: ashkan.shakarami@phd.unipd.it}, 
Azade Farshad\textsuperscript{2}, 
Yousef Yeganeh\textsuperscript{2},
Lorenzo Nicolè\textsuperscript{1},\\ 
Peter Schüffler\textsuperscript{2}, 
Stefano Ghidoni\textsuperscript{1}, 
Nassir Navab\textsuperscript{2}
}

\date{
\small \textsuperscript{1}University of Padova, Italy\\
\small \textsuperscript{2}Technical University of Munich, Germany\\[1ex]
\small \texttt{
ashkan.shakarami@phd.unipd.it, 
lorenzo.nicole@phd.unipd.it, 
stefano.ghidoni@unipd.it,\\
azade.farshad@tum.de, 
y.yeganeh@tum.de, 
peter.schueffler@tum.de, 
nassir.navab@tum.de
}
}

\begin{document}
\maketitle

\begin{abstract}
We propose \textbf{\method{}}, a unit-based tissue segmentation framework for histopathology that classifies each fixed-size \(32 \times 32\) tile, rather than each pixel, as the segmentation unit. This approach reduces annotation effort and improves computational efficiency without compromising accuracy. To implement this approach, we introduce a Multi-\textbf{L}evel \textbf{Vi}sion \textbf{T}ransformer (\net{}), which benefits the multi-level feature representation to capture both fine-grained morphology and global tissue context. Trained to segment breast tissue into three categories (\textit{infiltrating tumor}, \textit{non-neoplastic stroma}, and \textit{fat}), \method{} supports clinically relevant tasks such as tumor-stroma quantification and surgical margin assessment. Evaluated on 386{,}371 tiles from 459 H\&E-stained regions, it outperforms U-Net variants and transformer-based baselines. Code and Dataset will be available at \href{https://github.com/AshkanShakarami/UTS-P}{GitHub}.

\end{abstract}

\section{Introduction}

Accurate tissue segmentation in histopathological Whole Slide Images (WSIs) is fundamental to digital pathology, enabling computational support for diagnosis, prognosis, and treatment planning \cite{litjens2017survey, wan2020robust, Shakarami2025thesis}. Conventional approaches, such as U-Net \cite{ronneberger2015} and DeepLab \cite{chen2017deeplab}, rely on pixel-wise segmentation, which requires dense annotations, is prone to noisy outputs, and incurs substantial computational costs—especially at gigapixel resolution \cite{minaee2021, Xu2023}.

Transformer-based architectures such as TransUNet \cite{Chen2021}, Swin-UNet \cite{Cao2021}, and H2G-Net \cite{zhou2022h2gnet} have enhanced global context modeling and multiscale feature learning (A literature review is provided in \autoref{sec:literature_review}, which discusses related work in moew detail). However, they remain fundamentally tied to the pixel-wise paradigm, which limits scalability in clinical environments with hardware constraints.

We propose \textbf{\method{}}, which treats each fixed-size \(32 \times 32\) tile as a semantic unit rather than relying on conventional pixel-wise segmentation. This approach redefines the segmentation primitive to better align with how pathologists interpret tissue on WSI, and enables scalable analysis without the overhead of dense annotation. To operationalize this approach, we introduce \textbf{\net{}}, a new Multi-Level Vision Transformer architecture designed to capture local and global histological patterns.

Unlike pixel-wise segmentation at downsampled resolutions, which averages tissue information across regions, \method{} retains high-resolution morphological detail within each 32$\times$32 tile (\autoref{sec:Theoretical Comparison with Pixel-Wise}).

\net{} incorporates an EfficientNetB3 backbone and fuses multi-scale features through Multi-Level Feature Fusion (MLFF) for hierarchical representation learning. Attention modules, including Dilated Attention and Squeeze-and-Excitation (DAT-SE) and Dilated Convolutional Block Attention Module (D-CBAM), further enhance spatial and channel-level discrimination. By decoupling segmentation from pixel-level granularity and enabling rich context modeling, \method{} provides a computationally efficient, interpretable, and clinically aligned framework for whole-slide tissue analysis.

\section{\method{}}

\begin{figure*}
    \centering
    \includegraphics[angle=0, width=1\textwidth, trim={0cm, 0cm, 0cm, 0cm}, clip]{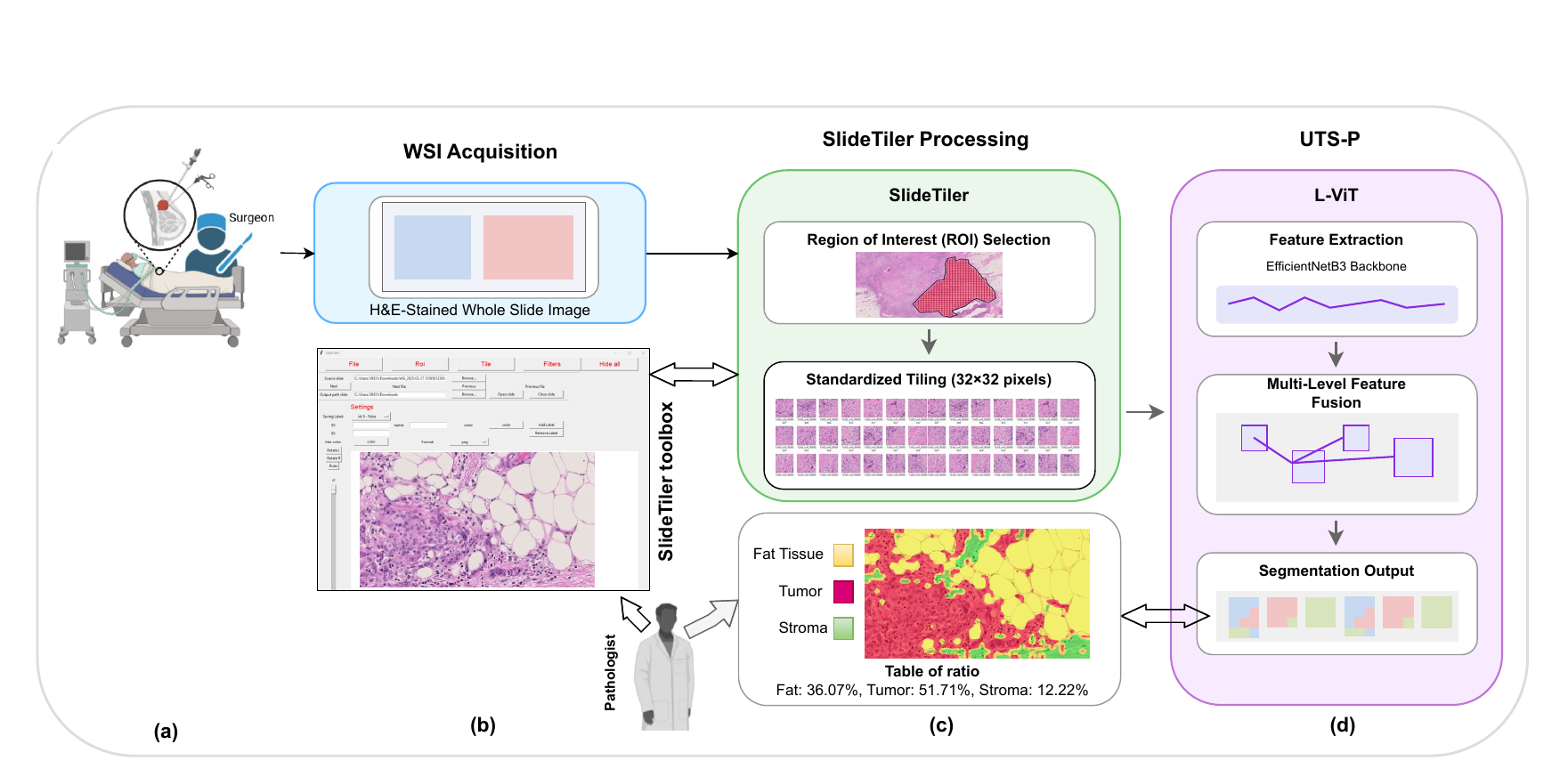}
    \captionsetup{justification=justified, singlelinecheck=false}
    \caption{\textbf{\method{}} \textbf{Pipeline}. \textbf{(a)} Tissue extraction via biopsy or surgery. \textbf{(b)} H\&E-stained WSI acquisition through processing, staining, and digitization. \textbf{(c)} SlideTiler preprocessing for WSI or ROI selection and standardized tiling into \(32 \times 32\) pixel tiles. \textbf{(d)} Segmentation using \net{} to classify tiles into \textit{Infiltrating Breast Tumor}, \textit{Fat Tissue}, and \textit{Non-neoplastic Stroma}, visualized with color-coded overlays. The system also computes tissue composition ratios (e.g., Tumor: 51.71\%, Stroma: 12.22\%, Fat: 36.07\%), enabling automated tumor-stroma quantification based on unit-level classification using fixed-size \(32 \times 32\) tiles. This integrated analysis supports Tumor–Stroma Ratio (TSR) estimation, providing interpretable metrics for prognostic assessment and personalized treatment planning.}

    \label{fig: pipeline}
\end{figure*}

\pipeline{}, as illustrated in \autoref{fig: pipeline}, presents an end-to-end framework for histopathological tissue segmentation, encompassing Data acquisition, preprocessing (\autoref{sec:WSI acquisition and preprocessing}), segmentation (\autoref{sec:segmentation}), and visualization (\autoref{subsec: segment_refinement}). \autoref{sec:Mathematical Justification} also proves theoretical justification of \method{}.

\subsection{Data Acquisition and Preprocessing}
\label{sec:WSI acquisition and preprocessing}

\method{} begins with the acquisition of H\&E-stained WSIs, digitized from tissue samples obtained through biopsy or surgical resection. To prepare these high-resolution WSIs for segmentation, we employ the SlideTiler toolbox \cite{Barcellona2024}, which facilitates efficient and standardized preprocessing.

In this stage, a pathologist selects either ROIs or entire slide areas. SlideTiler then partitions these selections into uniform \(32 \times 32\) pixel tiles. Formally, each WSI or ROI $I_i$ is divided into $N$ tiles $T_{ij}$, i.e., $I_i = \bigcup_{j=1}^{N} T_{ij}$, where each tile (semantic unit) $T_{ij}$ is stored in PNG format.

Unlike traditional tools such as QuPath, SlideTiler automates tile generation and preserves diagnostically relevant structures while excluding background and artifacts. This standardized tile-based format reduces preprocessing overhead and aligns with the unit-based segmentation approach of \method{}, allowing for seamless downstream processing by the \net{} segmentation engine (\autoref{sec:segmentation}). The resulting tiles serve as input to the classification model, forming a scalable and annotation-efficient representation of histological content. This tile-level granularity reflects how pathologists often interpret morphology in discrete regions, enabling efficient and interpretable segmentation aligned with clinical workflow.

\subsection{Segmentation}
\label{sec:segmentation}
The core segmentation engine of \method{} classifies each \(32 \times 32\) tile into one of three histological classes: \textit{infiltrating tumor}, \textit{non-neoplastic stroma}, or \textit{fat}. This classification approach is powered by \net{} (detailed in \autoref{subsec: L-ViT}), a vision transformer architecture adapted for histopathological tile segmentation. The model receives raw image tiles and outputs class probabilities directly, without requiring postprocessing or pixel-level masks. This simplifies annotation, reduces computational overhead, and aligns with how pathologists reason about tissue morphology at the regional level. By integrating the segmentation output over the entire slide or ROI, \method{} reconstructs a coarse-grained, semantically meaningful segmentation map that supports downstream clinical tasks such as tumor quantification and tissue composition analysis.

\subsubsection{\net{}}
\label{subsec: L-ViT}
\begin{figure}[h!]
    \centering
    \includegraphics[width=0.4\textwidth]{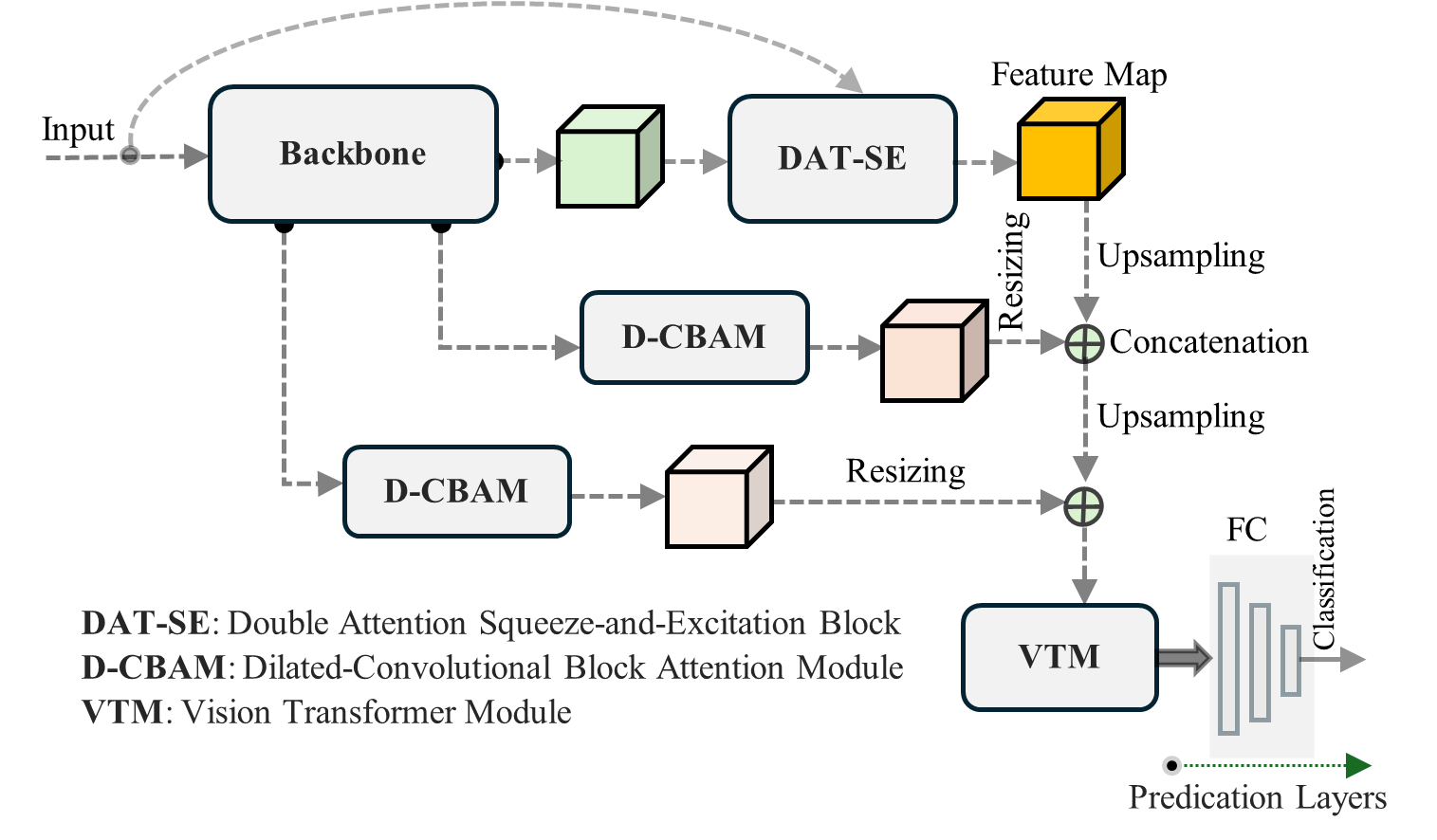}
    \caption{Overview of \net{}. MLFF integrates low-, mid-, and high-level features to enhance classification accuracy by combining local and global context. For component details, refer to DAT-SE (\autoref{subsec: DAT-SE}), D-CBAM (\autoref{subsec: D-CBAM}), and VTM (\autoref{subsec: VTM}).}
    \label{fig: L-ViT}
\end{figure}

\autoref{fig: L-ViT} illustrates an overview of \net{}. It adopts EfficientNetB3 as the backbone, taking advantage of its lightweight and efficient design, which offers high flexibility in adapting to varying input sizes and computational constraints through compound scaling \cite{Tan2019, Shakarami2023}. \net{} also integrates the DAT-SE for channel-wise feature recalibration (\autoref{subsec: DAT-SE}) \cite{hu2018} and D-CBAM to apply channel and spatial attention (\autoref{subsec: D-CBAM})\cite{woo2018}. In addition, MLFF \cite{Yang2019,Shakarami2021} ensures effective integration of features at different stages of the network. \net{} incorporates Transformer blocks in its final layers to capture long-range dependencies \cite{dosovitskiy2020}, as well. The refined feature maps are processed through dense layers, culminating in a softmax layer for multi-class classification. The following subsections provide further details on each component.

\begin{figure}[h!]
    \centering
    \includegraphics[width=0.4\textwidth]{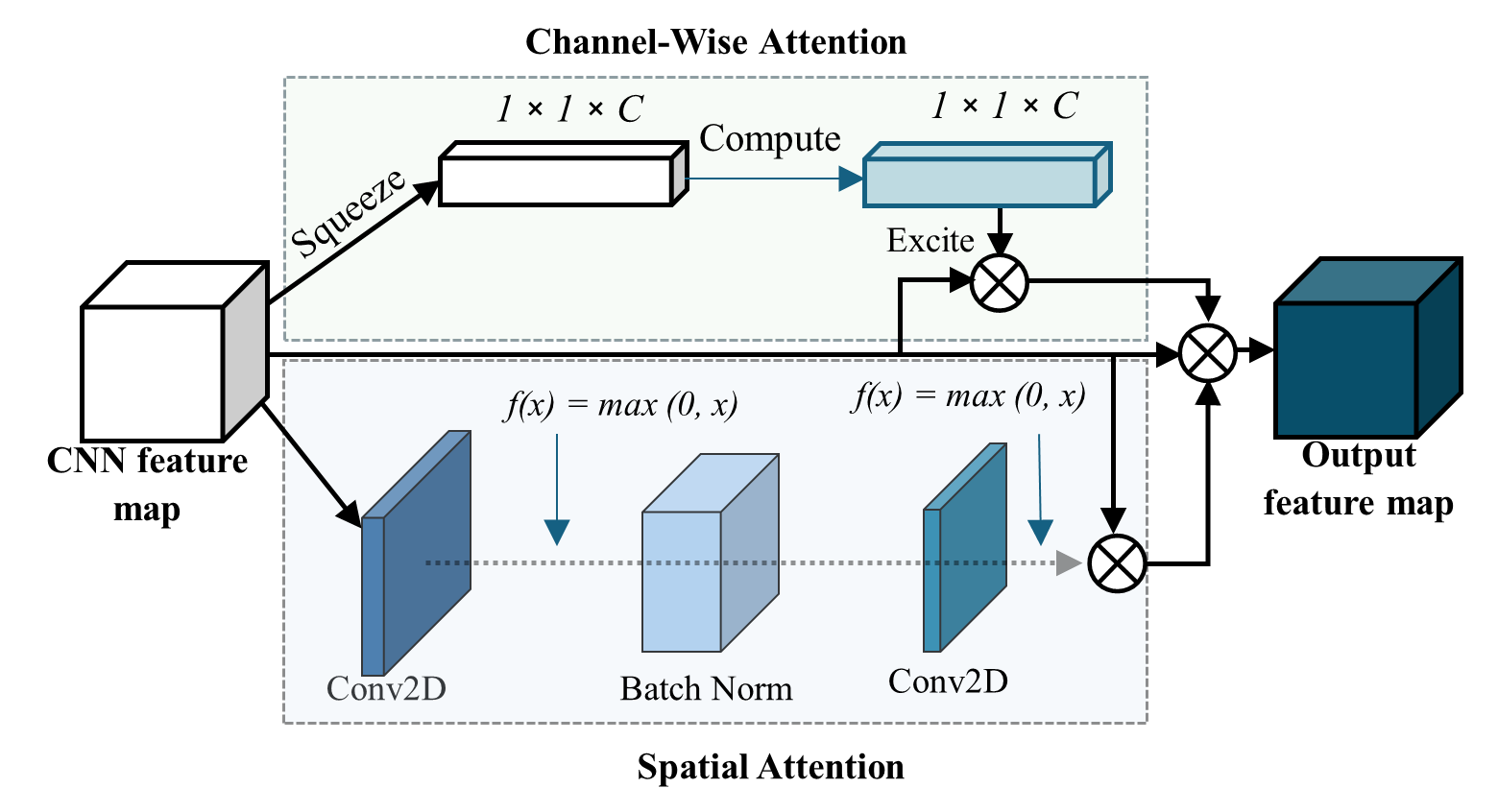}
    \caption{DAT-SE (\autoref{subsec: DAT-SE}): Enhances feature representation by recalibrating input maps with channel and spatial attention \cite{hu2018}.}
    \label{fig: DAT-SE}
\end{figure}

\begin{figure}[h!]
    \centering
    \includegraphics[width=0.4\textwidth]{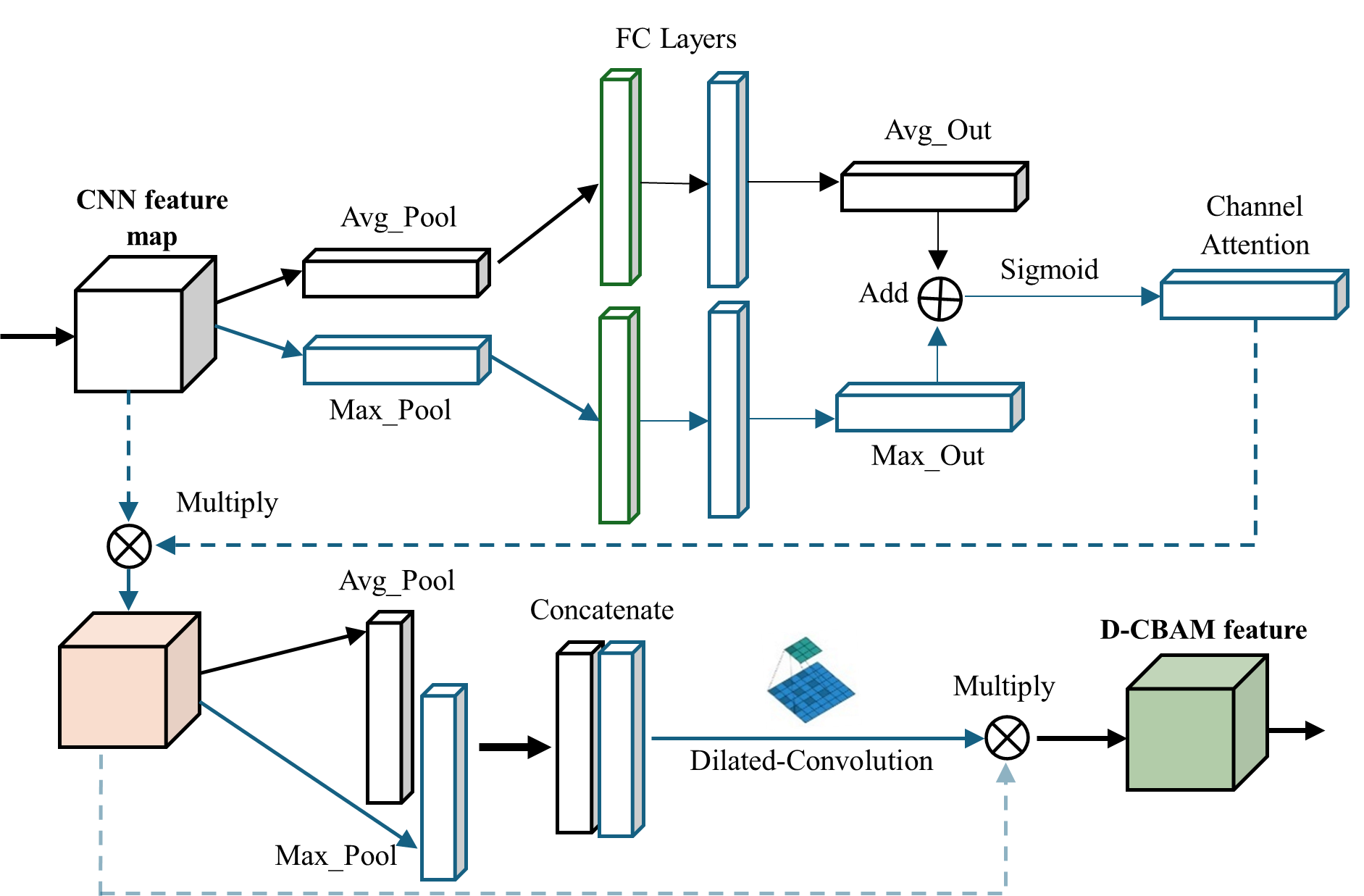}
    \caption{D-CBAM (\autoref{subsec: D-CBAM}): Extends CBAM \cite{Alirezazadeh2023, woo2018} by integrating channel and spatial attention for improved feature focus.}
    \label{fig: D-CBAM}
\end{figure}

\begin{figure}[h!]
    \centering
    \includegraphics[angle=0, width=0.49\textwidth]{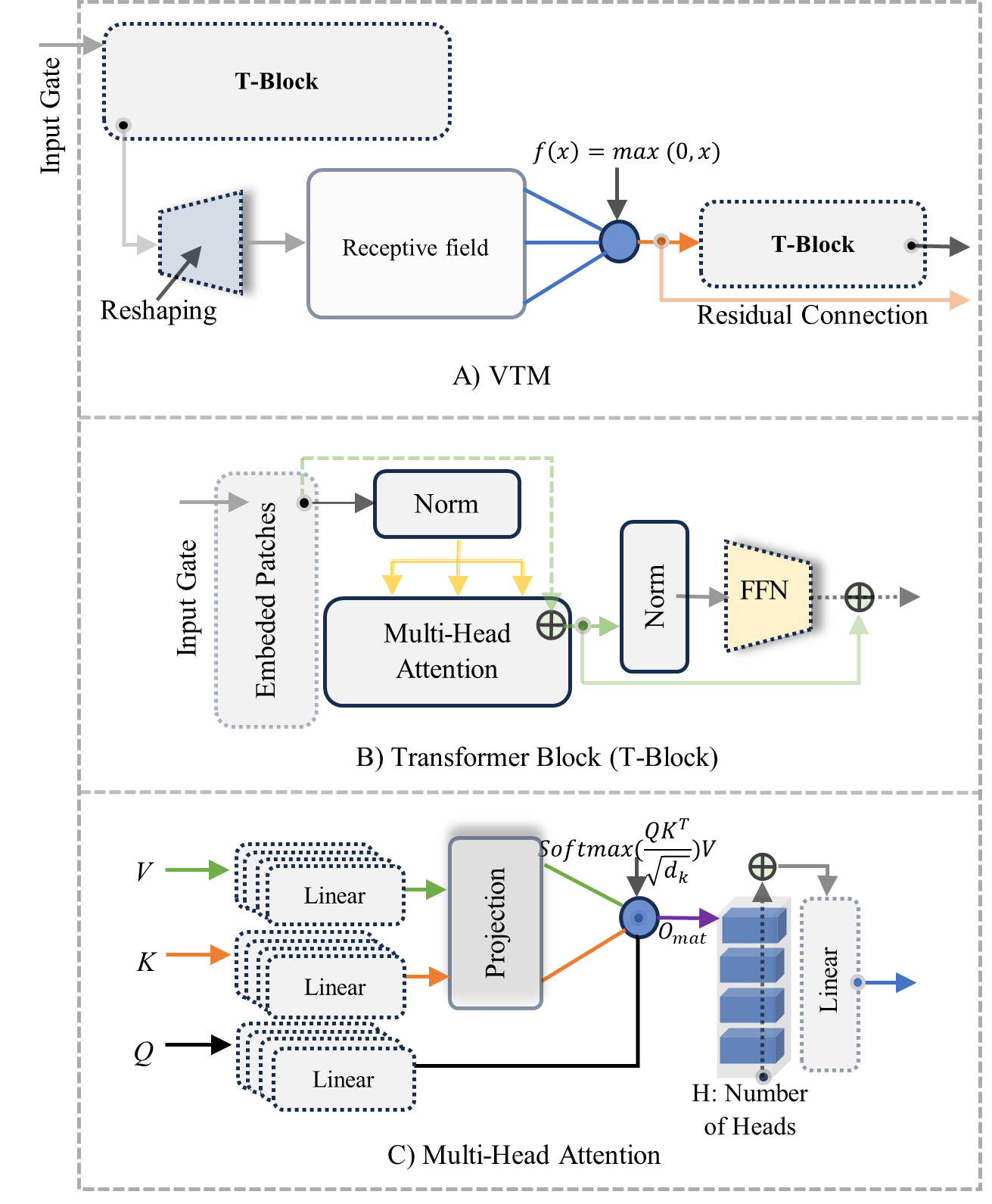}
    \caption{VTM Architecture (\autoref{subsec: VTM}), utilizing Transformer blocks \cite{dosovitskiy2020} with Multi-Head Self-Attention, Feed-Forward Networks, and Layer Normalization.}
    \label{fig:VTM}
\end{figure}

\paragraph{DAT-SE.}
\label{subsec: DAT-SE}
DAT-SE enhances feature representation by applying channel-wise and spatial attention mechanisms, dynamically recalibrating input feature maps to improve their discriminative power \cite{hu2018}. Given an input feature map \( \mathbf{X} \in \mathbb{R}^{H \times W \times C} \), channel attention is computed by extracting global context information through Global Average Pooling (GAP), transforming the feature map into a channel descriptor \( \mathbf{z} \in \mathbb{R}^{C} \). This representation is processed through two fully connected layers, followed by a sigmoid activation to generate channel-wise attention weights:

\begin{equation}
    \mathbf{s} = \sigma(\mathbf{W_2} \delta(\mathbf{W_1} \mathbf{z})).
    \label{eq:channel_attention_datse}
\end{equation}

Simultaneously, spatial attention enhances spatial feature relationships by capturing contextual dependencies. This is achieved through average and max pooling across the channel dimension, followed by a convolutional layer to generate spatial attention weights:

\begin{equation}
    \mathbf{M}_{\text{spatial}} = \sigma(\text{conv}([\text{avg}(\mathbf{X}); \text{max}(\mathbf{X})])).
    \label{eq:spatial_attention_datse}
\end{equation}

The final feature recalibration integrates both attention mechanisms, refining feature representation as:

\begin{equation}
    \mathbf{X}' = \mathbf{X} \cdot \mathbf{s} \cdot \mathbf{M}_{\text{spatial}}.
    \label{eq:final_refinement_datse}
\end{equation}

This dual-attention mechanism amplifies relevant features while suppressing irrelevant information, enhancing segmentation accuracy and robustness in histopathological image analysis.

\paragraph{D-CBAM.}
\label{subsec: D-CBAM}
D-CBAM is an improved Channel and Spatial Attention Mechanism inspired by CBAM \cite{Alirezazadeh2023, woo2018}. It enhances feature selection by integrating dilated convolutions and refining long-range feature dependencies while preserving fine-grained spatial details. Channel attention aggregates contextual information across channels using GAP and Global Max Pooling (GMP). These pooled features are then passed through fully connected layers, followed by sigmoid activation, to obtain channel-wise attention weights, formulated as 
\( \mathbf{M}_c = \sigma(\mathbf{W_0} (\delta(\mathbf{W_1} (\text{GAP}(\mathbf{X}))) 
+ \delta(\mathbf{W_1} (\text{GMP}(\mathbf{X}))))). \) Spatial attention enhances spatial feature representation by applying dilated convolutions over concatenated average and max pooled feature maps as \autoref{eq:spatial_attention_dcbam}.

\begin{equation}
    \mathbf{M}_s = \sigma(\text{conv}_{\text{dilated}}([\text{avgpool}(\mathbf{X}); \text{maxpool}(\mathbf{X})])).
    \label{eq:spatial_attention_dcbam}
\end{equation}

The final refined feature map is computed as \autoref{eq:final_refinement_dcbam}. D-CBAM effectively expands the receptive field by incorporating dilated convolutions, improving segmentation effectiveness by capturing local fine-grained and global structural information.

\begin{equation}
    \mathbf{X}' = \mathbf{X} \cdot \mathbf{M}_c \cdot \mathbf{M}_s.
    \label{eq:final_refinement_dcbam}
\end{equation}

\paragraph{VTM.}
\label{subsec: VTM}
VTM integrates Transformer blocks \cite{dosovitskiy2020} to model long-range dependencies and improve contextual understanding in histopathological tissue analysis. \autoref{fig:VTM} shows VTM with Transformer blocks (T-Block). Within each T-Block, the Multi-Head Self-Attention (MHSA) mechanism computes attention weights by processing the query, key, and value matrices, where each attention head is defined as \( \text{head}_i = \text{softmax} \left(\frac{Q_i K_i^T}{\sqrt{d_k}}\right) V_i \). The outputs from multiple heads are then concatenated to form the final multi-head attention output, formulated as \( \text{MHSA}(Q, K, V) = \text{concat}(\text{head}_1, ..., \text{head}_h) W_O \). Following MHSA, a FFN further refines feature representations through a two-layer transformation with ReLU activation, given by \( \text{FFN}(X) = \text{ReLU}(X W_1 + b_1) W_2 + b_2 \). To ensure stable training and effective feature transformation, Layer Normalization, and Residual Connections are applied, where the final normalized output is computed as \( Y_{\text{norm}} = \text{LayerNorm}(X + \text{MHSA}(Q, K, V) + \text{FFN}(X)) \). Moreover, VTM incorporates Multi-Head Linear Attention \cite{wang2020linformer}. This Linear Attention reduces the complexity from quadratic to linear with respect to the sequence length. This is achieved by projecting the query and key matrices into a lower-dimensional space, formulated as \( Q' = QW_Q, \quad K' = KW_K \).

\subsection{Visualization}
\label{subsec: segment_refinement}
\label{subsec: visualization}
The segmentation output obtained from the \method{} framework exhibits a tile-wise structure. As a result, edge boundaries are observed, which affects the spatial coherence of the segmented regions. Hence, we use the \textit{Segmentation Refinement} as a post-processing strategy with no additional computational cost (\autoref{subsec: refinement_complexity}), integrated end-to-end with \net{}, to refine the final segmentation output and address this issue. This strategy uses Neighborhood-Based Smoothing and Class Discretization, as explained below.

The refinement method consisted of two main stages: (\textit{i}) a local smoothing operation using a separable filter to mitigate abrupt transitions between neighboring tiles and (\textit{ii}) a class discretization step to enforce categorical consistency across the refined segmentation mask.

\paragraph{Neighborhood-Based Smoothing:}  
\label{subsec:Neighborhood-Based Smoothing}
A separable filtering approach was employed to refine the segmentation mask and enhance boundary consistency. This method computes the local mean of pixel values within a defined neighborhood of size \( 48 \times 48 \) pixels, corresponding to 1.5 times the tile size. Separable filtering decomposes the operation into two successive one-dimensional convolutions with a linear complexity (\( O(2k) \), where k denotes the neighborhood radius). The refined segmentation mask \( S_f(x, y) \) at a given pixel location \( (x, y) \) was computed in two sequential steps:

\vspace{8pt} 
\noindent
1. Horizontal Filtering: A one-dimensional convolution is first applied along the horizontal axis to smooth pixel intensities within a fixed neighborhood:
\begin{equation}
S'_f(x, y) = \frac{1}{N} \sum_{i=-k}^{k} S(x+i, y),
\label{eq:horizontal_smoothing}
\end{equation}

\noindent
2. Vertical Filtering: The intermediate result is then processed along the vertical axis to obtain the final refined segmentation mask:
\begin{equation}
S_f(x, y) = \frac{1}{N} \sum_{j=-k}^{k} S'_f(x, y+j),
\label{eq:vertical_smoothing}
\end{equation}

\noindent
where \( S(x, y) \) represents the original segmentation mask, \( k \) denotes the neighborhood radius, and \( N \) is the total number of pixels within the local region. This method achieves a substantial computational speedup by leveraging separable filtering while preserving segmentation precision, making it particularly suitable for large-scale histopathological image analysis.

\paragraph{Class Discretization:}  
Since the smoothing operation may introduce intermediate values deviating from the predefined class labels (absolute values for Red, Green, and Yellow representing Infiltrating Breast Tumors, Fat Tissue, and Non-neoplastic Stroma), a refinement step was applied to discretize the filtered output into distinct categories. Each pixel in the smoothed image was mapped to the closest predefined class by minimizing the Euclidean distance between its intensity vector and the corresponding class color centroid in the RGB space:

\begin{equation}
C(x, y) = \arg\min_{c \in \{R, G, Y\}} \| S_f(x, y) - C_c \|_2,
\end{equation}

where $C_c$ represents the class-specific color values for tumor (red: [255, 0, 0]), stroma (green: [0, 255, 0]), and fat (yellow: [255, 255, 0]). This step ensured the segmentation output remained interpretable, avoiding unsmooth edges introduced by continuous-valued tile-based filtering.

\begin{figure}[t]
    \centering
    \begin{subfigure}[b]{0.22\textwidth}
        \centering
        \includegraphics[width=\textwidth]{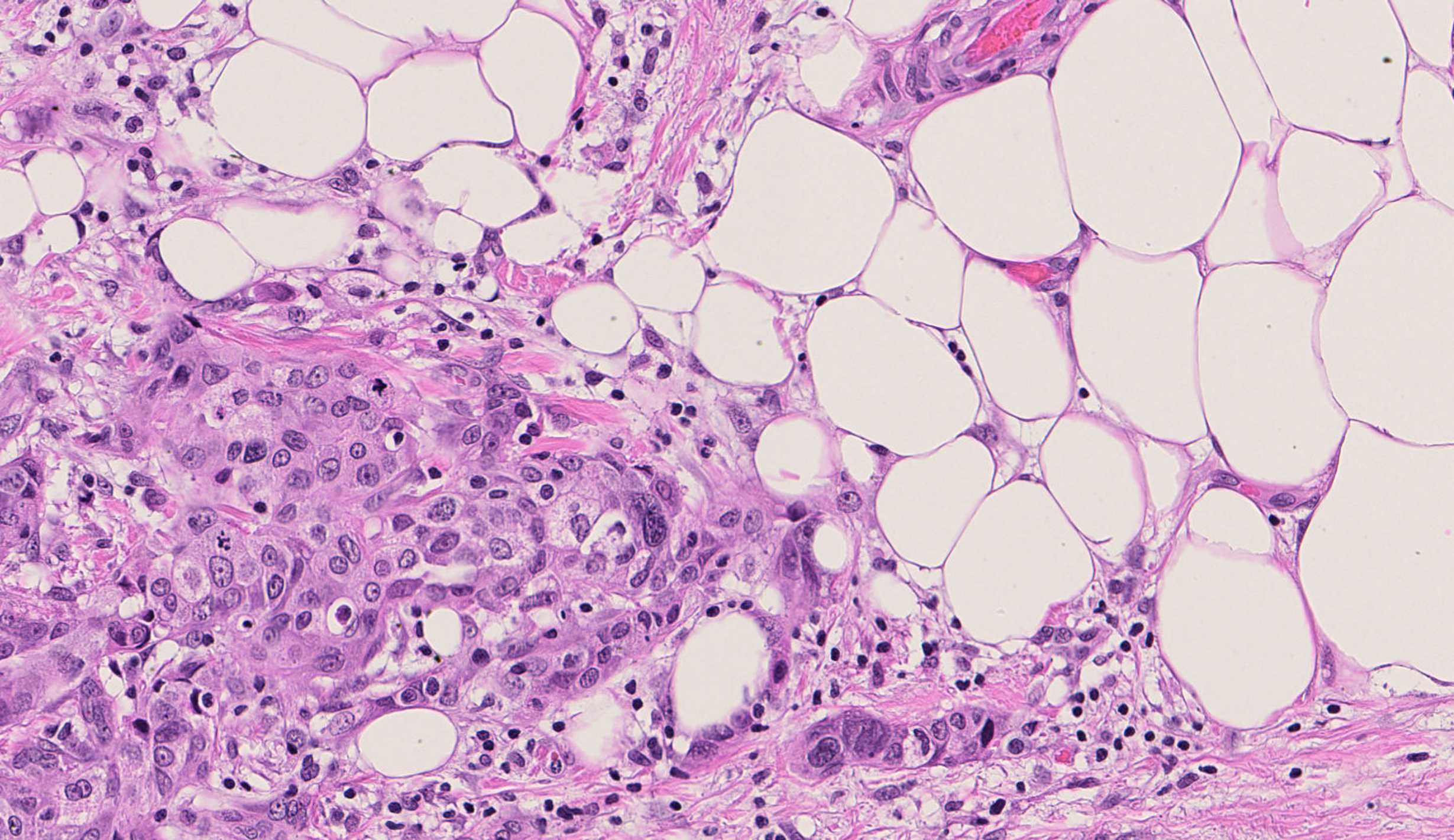}
        \caption*{(a)}
        \label{fig:original_roi}
    \end{subfigure}
    \hfill
    \begin{subfigure}[b]{0.22\textwidth}
        \centering
        \includegraphics[width=\textwidth]{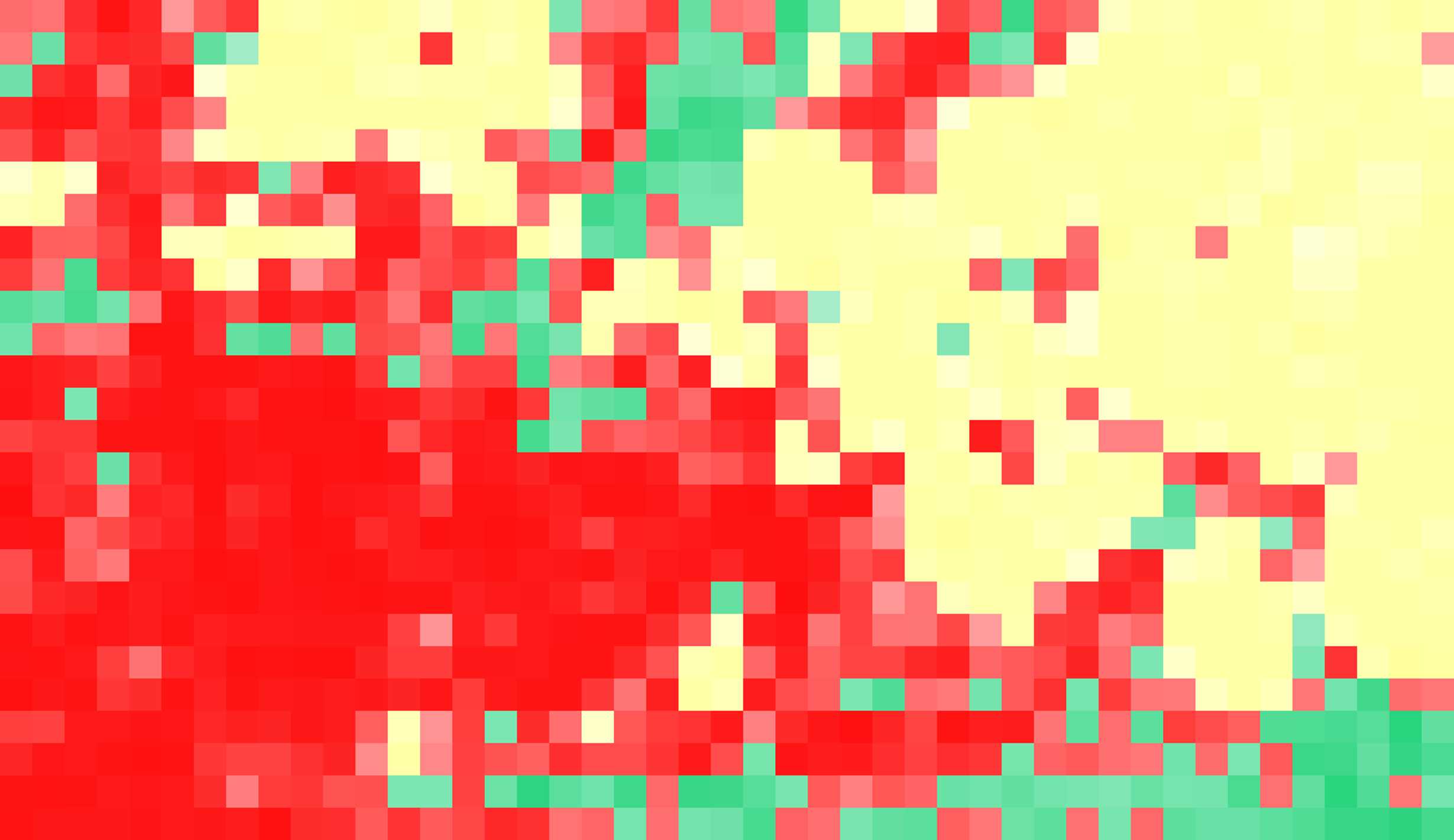}
        \caption*{(b)}
        \label{fig:tile_based_seg}
    \end{subfigure}

    \vspace{0.3cm}
    
    \begin{subfigure}[b]{0.22\textwidth}
        \centering
        \includegraphics[width=\textwidth]{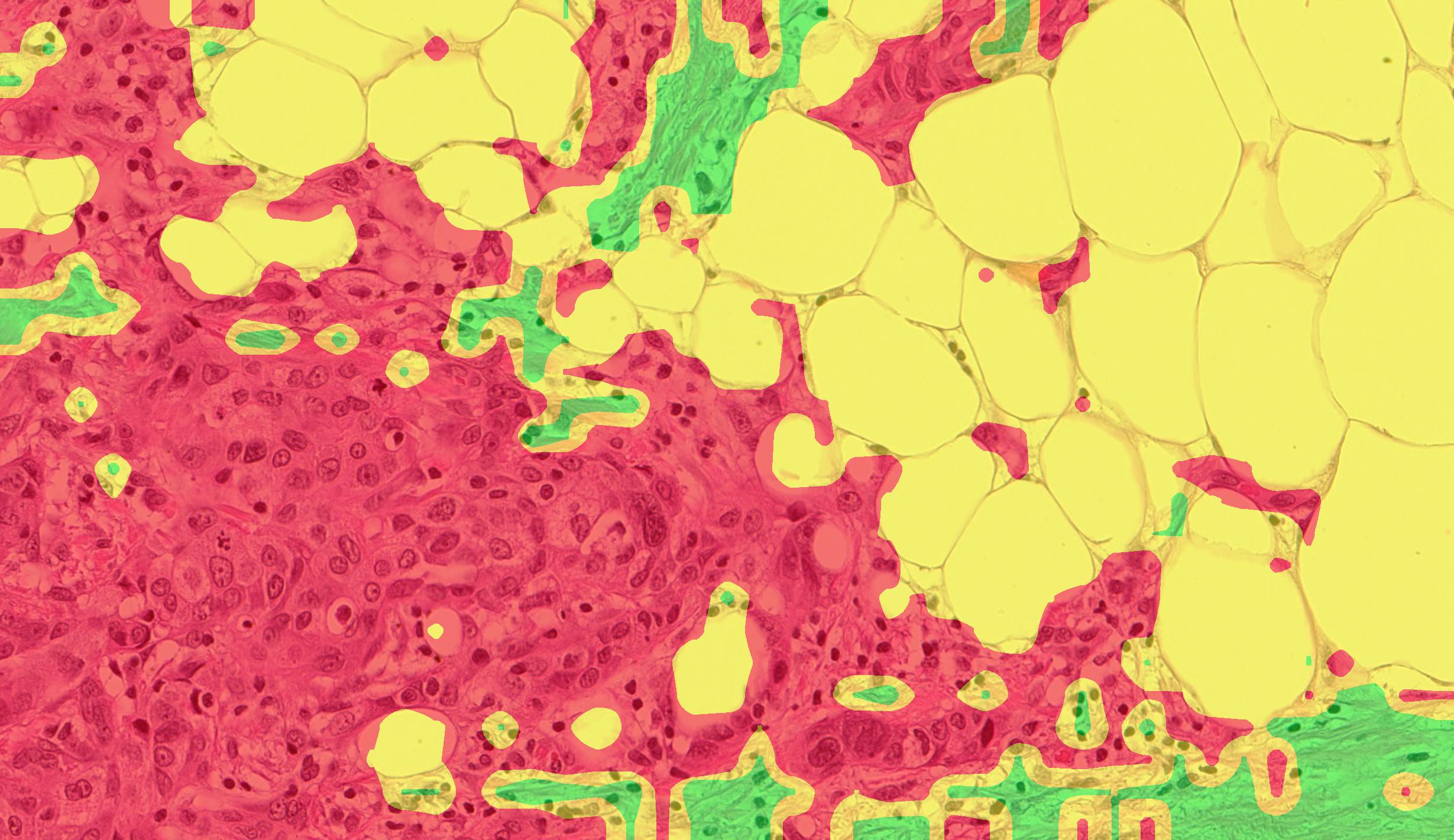}
        \caption*{(c)}
        \label{fig:overlay_seg}
    \end{subfigure}
    \hfill
    \begin{subfigure}[b]{0.22\textwidth}
        \centering
        \includegraphics[width=\textwidth]{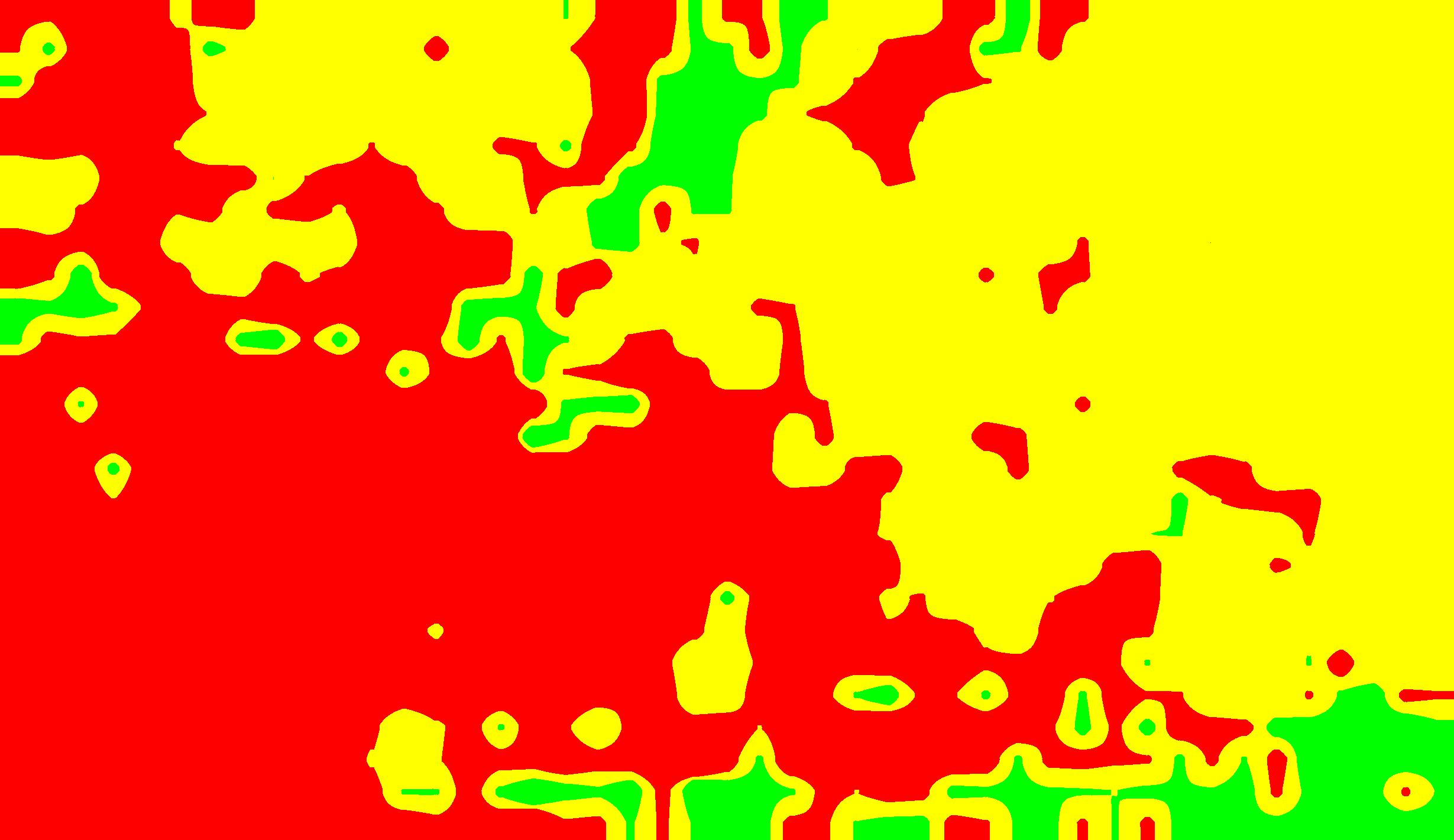}
        \caption*{(d)}
        \label{fig:final_seg}
    \end{subfigure}

    \caption{Visualization of segmentation refinement: 
    (a) Original ROI from H\&E-stained tissue,
    (b) unrefined segmentation output from tile classification,
    (c) refined segmentation overlaid on the original ROI, and 
    (d) final refined segmentation. Red, Yellow, and Green indicate \textit{Infiltrating Breast Tumors}, \textit{Fat Tissue}, and \textit{Non-neoplastic Stroma}, respectively.}
    \label{fig:segmentation_refinement}
\end{figure}

\paragraph{WSI Overlay:}  
Segmentation outputs should align with pathologists' diagnostic workflow to support clinical decision-making \cite{Chen2022, Shakarami2020a, Shakarami2021b, shakarami2024histo, Shakarami2023}. To support this, \method{} generates segmentation overlays that integrate with histological slides, preserving tissue morphology as shown in \autoref{fig:segmentation_refinement}. This enables direct validation of the method's predictions without disrupting standard diagnostic practices. \method{}'s visual outputs can function as an assistive tool rather than a replacement for expert judgment, allowing for direct comparison with histological evidence.

To visualize the refined segmentation within the context of the original histopathological slide, the final segmentation mask is applied to the original WSI using a weighted overlay technique:

\begin{equation}
I_{overlay} = \alpha S_d + (1 - \alpha) I_{WSI},
\end{equation}

where $S_d$ denotes the discretized segmentation, $I_{WSI}$ represents the original WSI, and $\alpha$ is a transparency coefficient ($\alpha = 0.5$). This visualization approach maintained the tissue morphology while clearly delineating the segmented regions.

\paragraph{Refinement Computational Complexity:}
\label{subsec: refinement_complexity}
The computational complexity of the segmentation refinement method is given by \(O(2k + 4\), where \( O(2k) \) arises from the separable filtering process (\autoref{subsec:Neighborhood-Based Smoothing}), which consists of two successive 1D convolutions (horizontal and vertical). \( O(3) \) corresponds to the class discretization step, where each pixel's RGB value is compared against the three predefined class colors. \( O(1) \) accounts for the WSI overlay, which applies a simple linear blending operation per pixel. Since \( k \) (the neighborhood radius) is larger than the constant terms, the overall computational complexity can be approximated as \( O(k) \), making the method highly efficient for large-scale histopathological image analysis.

In addition, the additional complexity above introduced by the segmentation refinement method, \(\mathcal{O}(k)\), is significantly lower than the existing complexities of the models in \autoref{tab:complexity_comparison}. Since the primary computational load in \net{} arises from tokenized tile-based segmentation, given by \(\mathcal{O}(M^2 \cdot D / k^2)\), adding a post-processing step with \(\mathcal{O}(k)\) complexity does not substantially impact the overall model complexity. as it is asymptotically negligible in comparison to \( \mathcal{O}(M^2 \cdot D / k^2) \). Therefore, the overall computational complexity of \net{} (\autoref{tab:complexity_comparison}) remains unchanged, and the dominant complexity term remains as \autoref{eq: refinement}.

\begin{equation}
\mathcal{O}(M^2 \cdot D / k^2) + \mathcal{O}(k) = \mathcal{O}(M^2 \cdot D / k^2).
\label{eq: refinement}
\end{equation}

\paragraph{Results and Impact:}
The refinement method improved segmentation smoothness and boundary continuity without compromising class accuracy. By integrating neighborhood-based filtering and class discretization, it effectively reduced tile-induced artifacts and enhanced clinical interpretability. As shown in \autoref{fig:segmentation_refinement}, subfigure (a) displays the original ROI, and (b) shows the raw tile-based segmentation output with visible blockiness. Subfigure (c) overlays the refined segmentation onto the ROI, highlighting improved spatial consistency, while (d) presents the final segmentation mask with smoother, more anatomically coherent class boundaries.

\section{Experiments}
\subsection{Dataset}
\label{subsec: dataset}
\label{subsec: pre-pro}
As shown in \autoref{tab:dataset}), the dataset comprises 459 ROIs, retrospectively selected from X anonymized H\&E-stained WSIs acquired at 20$\times$ magnification using the Ventana DP200 scanner. Each ROI was annotated by a board-certified pathologist to represent a single histological category—either \textit{Infiltrating Breast Tumor}, \textit{Non-neoplastic Stroma}, or \textit{Fat Tissue}—ensuring \textbf{purity} of class content within each ROI. These ROIs were then processed using SlideTiler \cite{Barcellona2024}, which partitions them into standardized 32$\times$32 pixel tiles. In total, 386{,}371 tiles were generated, distributed as follows: 100{,}221 fat tiles, 160{,}067 stroma tiles, and 126{,}083 tumor tiles. This class-pure ROI design eliminates ambiguous boundaries, simplifies annotation, and strengthens the reliability of class-specific learning during training. The size of the selected ROIs was between X and Y pixels.

\begin{table}
\centering
\caption{Dataset information (Number of Tiles and Samples per Category)}
\begin{adjustbox}{max width=0.45\textwidth}
\begin{tabular}{lcc}
\hline
\textbf{Categories} & \textbf{Tiles} & \textbf{ROIs} \\
\hline
Fat Tissue                             & 100221 & 153 \\
Non-neoplastic Stroma                  & 160067 & 153 \\
Infiltrating Breast Tumors             & 126083 & 153 \\
\hline
\textbf{Total}                         & \textbf{386371} & \textbf{459} \\
\hline
\end{tabular}
\end{adjustbox}
\label{tab:dataset}
\end{table}

\subsection{Performance}
\label{sec:HistoCancer_Prediction_Performance}

All reported performance metrics are computed as \textbf{macro-averages} across the three tissue classes to ensure equal contribution from each category, regardless of class imbalance.

\method{} outperformed baseline CNNs (MobileNetV2~\cite{Sandler2018}, ResNet50~\cite{He2016}, and EfficientNet-B3~\cite{Tan2019}) as well as state-of-the-art models such as Attention U-Net~\cite{Oktay2018}, ResUNet++~\cite{Zhang2020}, and H2G-Net~\cite{Zhou2022}, as shown in \autoref{tab:results_baseline} and \autoref{tab:results_SOTA}. A 3-fold patient-level cross-validation ensured robust evaluation and avoided data leakage.

Compared to MobileNetV2 (IoU: \(72.68 \pm 6.43\%\), DSC: \(83.76 \pm 4.92\%\)), \method{} achieved \(84.44 \pm 1.89\%\) IoU and \(91.37 \pm 1.64\%\) DSC, reflecting gains of \(11.76\%\) and \(7.61\%\), respectively. Against ResNet50 (IoU: \(80.25 \pm 3.06\%\), DSC: \(88.95 \pm 2.74\%\)), improvements of \(4.19\%\) in IoU and \(2.42\%\) in DSC were observed. With approximately 128{,}790 tiles per validation fold (\autoref{tab:dataset}), each 1\% improvement translates to over 1{,}200 additional correctly segmented tiles—highlighting the clinical relevance of even marginal gains.

Although \net{} yielded the best results, one of \method{}'s notable advantages is its architectural flexibility. CNNs like ResNet50 still achieved high performance within the unit-based segmentation pipeline, benefiting from the structured tile-level approach. This demonstrates that \method{} is a robust host framework adaptable to both lightweight and high-capacity models, enabling deployment across a range of clinical and hardware environments.

\begin{table}[ht]
    \centering
    \caption{Comparison of \net{} with CNNs within \method{} framework}
    \label{tab:results_baseline}
    \scalebox{0.7}{
    \begin{tabular}{l c c c c c}
        \toprule
        Model & Acc & R & Spec & DSC & IoU \\
        \midrule
        MobileNetV2 \cite{Sandler2018} & 84.76 & 81.60 & 90.36 & 83.76 & 72.68 \\
    
        ResNet50 \cite{He2016} & 88.79 & 89.27 & 93.83 & 88.95 & 80.25 \\
   
        EfficientNet-B3 \cite{Tan2019} & 89.98 & 89.41 & 94.15 & 89.91 & 81.74 \\
      
        \textbf{\net{} } & \textbf{90.92} & \textbf{91.06} & \textbf{95.10} & \textbf{91.37} & \textbf{84.44} \\
        \bottomrule
    \end{tabular}
    }
\end{table}

Beyond CNN-based models, \method{} was benchmarked against advanced architectures such as Attention U-Net, ResUNet++, H2G-Net, and BEFUnet. BEFUnet, the most competitive state-of-the-art model, achieved an IoU of \(77.60\%\) and DSC of \(87.10\%\) (\autoref{tab:results_SOTA}). However, \method{} outperformed all models, achieving the highest DSC (\(91.37\%\)) and IoU (\(84.44\%\)). These improvements demonstrate \method{}’s superior capability in differentiating between histological components, including Infiltrating Breast Tumor, Fat Tissue, and Non-neoplastic Stroma regions.

\begin{table}[h!]
    \centering
    \caption{Comparison of \method{} with State-of-the-Art Models}
    \label{tab:results_SOTA}
    \small
    \scalebox{0.7}{
    \begin{tabular}{l c c c c c}
        \toprule
        Model & Acc & R & P & DSC & IoU \\
        \midrule
        Attention U-Net \cite{Oktay2018} & - & 84.00 & 82.50 & - & - \\
        ResUNet++ \cite{Zhang2020} & - & 70.22 & 87.85 & - & 79.62 \\
        H2G-Net \cite{Zhou2022} & - & 87.90 & 90.70 & - & - \\
        BEFUnet \cite{BEFUnet2024} & 87.03 & 90.80 & 87.30 & 87.10 & 77.60 \\
        \textbf{\method{} (Ours)} & \textbf{90.92} & \textbf{91.06} & \textbf{91.68} & \textbf{91.37} & \textbf{84.44} \\
        \bottomrule
    \end{tabular}
    }
\end{table}

\subsection{Ablation Study}
To evaluate the contributions of individual components in our proposed \textbf{\net{}} architecture, we conducted an ablation study by incrementally incorporating key modules and analyzing their effects on segmentation performance. ~\autoref{tab:ablation} summarizes the results of different architectural configurations in terms of DSC, Recall, IoU, Accuracy, and Specificity.

\paragraph{Baseline Model.}
The baseline model utilizes EfficientNet-B3 as the backbone without additional enhancement modules. As shown in ~\autoref{tab:ablation}, this configuration achieves a DSC of $89.91 \pm 0.54$, an IoU of $81.74 \pm 0.75$, and an accuracy of $89.98 \pm 0.56\%$. While EfficientNet-B3 provides strong feature extraction, it lacks advanced spatial and attention-based feature refinement, which limits overall segmentation accuracy.

\paragraph{Effect of VTM.}
Integrating the VTM with EfficientNet-B3 improves performance across all metrics. The DSC increases to $90.67$, and IoU improves to $83.36$, indicating that the VTM effectively captures long-range dependencies and enhances feature representation. This demonstrates that replacing purely convolutional feature extraction with transformer-based modeling yields better segmentation outcomes. As previously noted, even slight performance improvements can noticeably enhance segmentation performance. For instance, a 1\% increase in segmentation performance translates to 1287 additional correctly segmented tiles during evaluation (Given the scale of our dataset (\autoref{tab:dataset}), with approximately 128790 tiles per validation fold, such gains can have a substantial impact).

\paragraph{Contribution of DAT-SE.}
Further incorporating DAT-SE alongside VTM enhances feature recalibration by dynamically adjusting channel-wise importance. This further increases DSC ($90.85$) and IoU ($83.63$), with recall reaching $91.16\%$.

\paragraph{Impact of MLFF and D-CBAM.}
By jointly incorporating MLFF and D-CBAM, \net{} reaches the highest segmentation performance: DSC: $91.37 \pm 1.13$, IoU: $84.44 \pm 1.89$, and Accuracy: $90.92 \pm 1.14$. These improvements confirm that MLFF effectively enhances information propagation across different feature levels, while D-CBAM refines spatially important regions, leading to a more robust segmentation performance.

\begin{table}[h!]
\centering
\caption{Ablation study results for different \net{} components.}
\label{tab:ablation}
\scalebox{0.7}{
\begin{tabular}{l c c c c c}
\toprule
\textbf{Component} & \textbf{DSC} & \textbf{R} & \textbf{IoU} & \textbf{Acc} & \textbf{Spec} \\
\midrule
Backbone & 89.91 & 89.41 & 81.74 & 89.98 & 94.15 \\
+ VTM & 90.67 & 91.11 & 83.36 & 90.68 & 95.00 \\
+ VTM \& DAT-SE & 90.85 & 91.16 & 83.63 & 90.68 & 95.05 \\
\textbf{All (\net{})} & \textbf{91.37} & \textbf{91.06} & \textbf{84.44} & \textbf{90.92} & \textbf{95.10} \\
\bottomrule
\end{tabular}
}
\end{table}

As the final observation, the ablation study confirms that each component in \net{} contributes to segmentation quality. The backbone (EfficientNet-B3) provides strong initial feature extraction, VTM introduces global attention, DAT-SE refines feature representation, MLFF ensures hierarchical feature integration, and D-CBAM further improves spatial attention. 

\subsection{Efficiency and Advantages}
\label{subsec:computational_efficiency}

Traditional pixel-wise segmentation in WSI analysis requires up to \(N = 512 \times 512 = 262{,}144\) operations per patch, making it computationally intensive. In contrast, \method{} reduces the burden by processing only 256 non-overlapping tiles per image when using \(32 \times 32\) units—yielding a 1024-fold reduction in operations per patch while maintaining segmentation accuracy.

Compared to CNN-based architectures such as U-Net~\cite{ronneberger2015}, with complexity \(\mathcal{O}(N \cdot K^2 \cdot C_{\text{in}} \cdot C_{\text{out}})\), \method{} benefits from a simplified input structure and a more scalable processing path. The use of tokenized inputs and multi-level fusion in \net{} reduces computational complexity to \(\mathcal{O}(M^2 \cdot D / k^2)\), where \(k\) is the tile size. The lightweight refinement step, discussed in \autoref{subsec: segment_refinement}, adds only linear-time complexity \(\mathcal{O}(k)\) and does not affect overall scalability. This efficiency translates well to real-world conditions. We benchmarked \method{} on a single NVIDIA GeForce RTX 3060 GPU with an Intel® Core™ i7-11800U CPU and 16 GB RAM. Training required only 90 seconds per epoch (batch size 64), demonstrating the framework’s compatibility with modest hardware resources~\cite{Shen2022efficient, Wang2021annotation}.

\begin{table}[h!]
    \centering
    \caption{Computational Complexity Comparison}
    \label{tab:complexity_comparison}
    \scalebox{0.7}{
    \begin{tabular}{l c c}
        \toprule
        \textbf{Model} & \textbf{Complexity} & \textbf{Approach} \\
        \midrule
        U-Net \cite{ronneberger2015} & $\mathcal{O}(N \cdot K^2 \cdot C_{\text{in}} \cdot C_{\text{out}})$ & Pixel-wise \\
        HIPT \cite{Chen2022} & $\mathcal{O}(M^2 \cdot D)$ & Patch-based \\
        \textbf{\method{}} & $\mathcal{O}(M^2 \cdot D / k^2)$, $k$: tile dimension & Unit-based \\
        \bottomrule
    \end{tabular}
    }
\end{table}

In addition to efficiency, \method{} offers practical advantages through its design. The integration of MLFF consolidates low- and high-level features, enhancing robustness across heterogeneous histological regions. Coupled with the EfficientNet-B3 backbone, this enables context-aware segmentation that balances resolution and generalizability. \method{} also supports interpretability by producing decision maps and probability-based heatmaps that can be directly overlaid onto WSIs, an essential feature for diagnostic workflows involving uncertainty and tissue complexity.

\subsection{1$\times$ vs. 20$\times$ Resolution}
\label{sec:Theoretical Comparison with Pixel-Wise}
Although segmentation via classification is not novel per se, \method{} offers a key advantage by operating on full-resolution 32$\times$32 tiles, which preserve fine-grained morphological features essential for histopathological analysis. In contrast, a pixel-wise segmentation model applied to 1$\times$ downsampled WSIs would treat each pixel as representing an entire 32$\times$32 patch from the original resolution. This downsampling process effectively averages information over those regions, resulting in the loss of fine structural details such as nuclear boundaries, gland formations, or stromal texture.

Therefore, while a 1$\times$ pixel-wise segmentation may seem equivalent in spatial scale to unit-based classification, it lacks access to the high-resolution content. \method{}, by classifying tiles in the standard resolution (20$\times$), retains diagnostic features that are crucial for accurate tissue discrimination. This theoretical advantage aligns with pathologists' reliance on high-resolution morphology and supports \method{}’s superior performance in segmenting tumor, stroma, and fat tissue with high fidelity.

\section{Limitations and Future Work}

While \method{} demonstrates strong segmentation performance, several limitations remain. Blank or non-informative regions in WSIs, such as background or slide borders that do not belong to tumor, stroma, or fat classes, were excluded from the dataset to avoid label noise and inconsistent supervision. While this improves training stability, it limits direct generalization to whole slides with such regions, and future work may introduce an explicit null-class for handling them. Moreover, \method{} currently employs a custom transformer trained from scratch; integrating foundation models like DINOv2~\cite{Oquab2023Dinov2} could improve generalization and reduce training cost. The use of fixed \(32 \times 32\) tiles also restricts compatibility with datasets such as CAMELYON16~\cite{Camelyon16Data}, which require tiling standardization. In addition, MLFF uses fixed-weight fusion, which could be improved with adaptive strategies to better capture intra-slide variability. Finally, \method{} may serve as a fast front-end module for coarse-to-fine pipelines~\cite{Wang2025CFI-ViT}.

Although \method{} currently focuses on segmenting three primary breast tissue types (tumor, stroma, and fat), it can be readily extended to additional categories such as necrosis, lymphocytic infiltration, or benign epithelium, provided appropriate annotations. Moreover, while our experiments were conducted on expert-defined ROIs to ensure purity and class balance, \method{} is architecturally compatible with whole-slide inference. Given its tile-based design and runtime efficiency (see \autoref{subsec:computational_efficiency}), full-slide segmentation is feasible through parallelized batch inference and will be investigated in future large-scale deployments.

\section{Conclusion}
We introduced \textbf{UTS-P}, a unit-based segmentation framework that redefines the segmentation primitive from pixels to tiles, enabling scalable and interpretable histopathological image analysis. Using a lightweight yet powerful architecture (\textbf{L-ViT}), integrating multi-level feature fusion, hierarchical attention modules, and a refinement strategy, \method{} achieves high segmentation performance on clinically relevant breast tissue categories: \textit{infiltrating tumor}, \textit{non-neoplastic stroma}, and \textit{fat}. Our results demonstrate that unit-based segmentation can match or exceed the performance of state-of-the-art pixel-wise methods while offering improvements in computational efficiency, annotation cost, and visual interpretability. Furthermore, \method{}'s compatibility with resource-constrained hardware makes it a practical solution for real-world clinical environments. We believe this approach opens new avenues for scalable, annotation-efficient, and diagnostically aligned segmentation systems in digital pathology and beyond.

\small
\bibliographystyle{unsrt}
\bibliography{references}

\appendix

\section{Theorem}
\label{sec:Mathematical Justification}
This section provides a theoretical foundation for the unit-based segmentation approach in \method{}, where each tile of size \(k \times k\) serves as the fundamental unit of prediction. The formulation addresses key challenges in histopathological image segmentation, namely label noise, computational complexity, and generalization in high-dimensional spaces.

\paragraph{Variance Reduction via Tile Aggregation.}
Let \(\mathcal{X}\) denote the space of WSIs, \(\mathcal{Y} = \{1, \dots, C\}\) the label space, and \(x_p \in \mathcal{X}_p\) a pixel from a WSI. In pixel-wise segmentation, a model \(f: \mathcal{X}_p \rightarrow \mathcal{Y}\) may suffer from annotation noise, especially at decision boundaries or under staining variability. Let \(x_t\) be a tile containing \(k^2\) pixels, then its label \(y_t\) is assigned via majority voting. The variance of the label noise is reduced as:

\[
\mathbb{V}[y_t] \leq \frac{1}{k^2} \sum_{i=1}^{k^2} \mathbb{V}[y_i],
\]

assuming independence across pixel labels. This makes tile-level predictions inherently more stable and less sensitive to noise compared to pixel-wise classification.

\paragraph{Multi-Level Feature Representation.}
Let \(\phi_l, \phi_m, \phi_h\) denote features extracted at low, mid, and high abstraction levels within the encoder. MLFF module aggregates these as:

\[
f_{\text{MLFF}} = \phi(\phi_l + \phi_m + \phi_h),
\]

where \(\phi(\cdot)\) is a nonlinear transformation. This can be interpreted as a residual composition over multiscale spaces. Theoretical results from function approximation show that multi-resolution architectures yield higher representational capacity: \( \mathcal{F}_{\text{MLFF}} \supset \mathcal{F}_{\text{shallow}} \), enabling better convergence on complex tissue boundaries.

\paragraph{Attention Mechanisms as Input-Adaptive Kernels.}
Channel and spatial attention modules (DAT-SE and D-CBAM) act as dynamic filters. Given an input \(x\), the attention-weighted output is:

\[
f_{\text{att}}(x) = A(x) \odot f(x),
\]

where \(A(x)\) is an attention map and \(\odot\) denotes element-wise multiplication. This modulates feature activations conditioned on input statistics, mimicking adaptive kernel selection:

\[
K_{\text{att}}(x,x') = \langle A(x)f(x), A(x')f(x') \rangle,
\]

which improves class separation and supports context-dependent decision-making.

\paragraph{Bias–Variance Trade-off in High Dimensions.}
For the generalization error \( \mathcal{E} = \text{Bias}^2 + \text{Variance} + \sigma^2 \), where \(\sigma^2\) denotes irreducible error, \method{} reduces both bias and variance through architectural design:

- \textbf{Variance reduction} from tile aggregation and MLFF,
- \textbf{Bias reduction} from attention-based adaptive modulation.

Thus, \method{} improves learning efficiency in limited-label regimes and stabilizes optimization in high-dimensional feature spaces.

\paragraph{Complexity and Scalability.}
Pixel-wise models like U-Net have per-image complexity \(\mathcal{O}(N \cdot K^2)\), where \(N\) is the number of pixels and \(K\) the kernel size. In contrast, unit-based segmentation performs \(\mathcal{O}(M^2 \cdot D / k^2)\) operations for \(M \times M\) token sequences and tile size \(k\), yielding a significant speedup while preserving resolution-aware segmentation. This aligns with the practical need for scalable inference on gigapixel WSIs.

\section{Literature Review}
\label{sec:literature_review}

Tissue segmentation in histopathological WSIs plays a central role in computational pathology, enabling high-throughput diagnostic support and quantitative tissue analysis~\cite{litjens2017survey,minaee2021image}. Traditional convolutional architectures such as U-Net~\cite{ronneberger2015u} and DeepLab~\cite{chen2017deeplab} perform dense pixel-wise segmentation, leveraging local receptive fields and spatial skip connections. Despite their effectiveness, these models are computationally intensive and demand densely annotated masks.

To address the limitations of locality and scalability, recent segmentation models have integrated transformers. Architectures like TransUNet~\cite{chen2021transunet}, Swin-UNet~\cite{cao2021swin}, and H2G-Net~\cite{zhou2022h2gnet} extend CNN backbones with self-attention to model long-range dependencies. Enhanced models such as BEFUnet~\cite{wang2024befunet} and HistoFormer~\cite{sun2024histogram} incorporate specialized modules for boundary enhancement and hierarchical context reasoning. Although effective in increasing spatial awareness, these models still rely on pixel-level granularity, which can introduce noise and hinder interpretability in clinical settings.

An emerging direction addresses the annotation bottleneck by shifting from pixel-wise to unit-based modeling, where each fixed-size tile is treated as a semantic entity. This design aligns with the diagnostic reasoning of pathologists, who often assess tissue in coherent regions rather than at the pixel level. Prior studies in weakly supervised learning and patch-level classification have laid the groundwork for this paradigm, but often lack structural consistency or systematic integration into segmentation pipelines.

Our work advances this research line by formalizing a unit-based framework, \method{}, which segments WSIs at the tile level using a dedicated transformer backbone (\method{}). Through multi-level feature fusion and adaptive attention mechanisms such as DAT-SE~\cite{hu2018squeeze} and D-CBAM~\cite{woo2018cbam}, the framework combines local morphology with global tissue context, providing a scalable and annotation-efficient solution for whole-slide histological analysis.

\end{document}